\documentclass[pre,amsmath,amssymb,aps,reprint,groupedaddress,nofootinbib]{revtex4-2} 
\pdfoutput=1
\usepackage[T1]{fontenc}
\usepackage[utf8]{inputenc}
\usepackage{lmodern}
\usepackage{pgf}
\usepackage{tikz}
\usepackage{booktabs}
\usepackage{scalerel}

\DeclareUnicodeCharacter{2212}{\ensuremath{-}}

\usetikzlibrary{svg.path}
\definecolor{orcidlogocol}{HTML}{A6CE39}
\tikzset{
  orcidlogo/.pic={
    \fill[orcidlogocol] svg{M256,128c0,70.7-57.3,128-128,128C57.3,256,0,198.7,0,128C0,57.3,57.3,0,128,0C198.7,0,256,57.3,256,128z}; 
    \fill[white] svg{M86.3,186.2H70.9V79.1h15.4v48.4V186.2z}
                 svg{M108.9,79.1h41.6c39.6,0,57,28.3,57,53.6c0,27.5-21.5,53.6-56.8,53.6h-41.8V79.1z M124.3,172.4h24.5c34.9,0,42.9-26.5,42.9-39.7c0-21.5-13.7-39.7-43.7-39.7h-23.7V172.4z} 
                 svg{M88.7,56.8c0,5.5-4.5,10.1-10.1,10.1c-5.6,0-10.1-4.6-10.1-10.1c0-5.6,4.5-10.1,10.1-10.1C84.2,46.7,88.7,51.3,88.7,56.8z}; 
  }
}

\newcommand\orcidicon[1]{\href{https://orcid.org/#1}{\mbox{\scalerel*{
\begin{tikzpicture}[yscale=-1,transform shape]
\pic{orcidlogo};
\end{tikzpicture}
}{D}}}}

\usepackage[citecolor=blue,colorlinks=true,pdfusetitle,
pdfauthor={Vishnu V. Krishnan, Kabir Ramola, Smarajit Karmakar},
pdfsubject={Statistical Physics of Disordered systems},
pdfkeywords={Glass, Hessian, Eigenvalue, Isotropy}]{hyperref}

\begin{document}

\title{Universal non-Debye low-frequency vibrations in sheared amorphous solids}

\author{Vishnu \surname{V. Krishnan}\thinspace\orcidicon{0000-0003-3889-3214}} 
\email{vishnuvk@tifrh.res.in}
\author{Kabir Ramola\thinspace\orcidicon{0000-0003-2299-6219}} 
\email{kramola@tifrh.res.in}
\author{Smarajit Karmakar\thinspace\orcidicon{0000-0002-5653-6328}} 
\email{smarajit@tifrh.res.in}
\affiliation{Centre for Interdisciplinary Sciences, Tata Institute of
Fundamental Research, Hyderabad 500046, India}

\date{\today}

\begin{abstract}
    We study energy minimized configurations of amorphous solids with a simple shear degree of freedom. We show that the low-frequency regime of the vibrational density of states of structural glass formers is crucially sensitive to the stress-ensemble from which the configurations are sampled. In both two and three dimensions, a shear-stabilized ensemble displays a $D(\omega_{\min}) \sim \omega^{5}_{\min}$ regime, as opposed to the $\omega^{4}_{\min}$ regime observed under unstrained conditions. We also study an ensemble of two dimensional, strained amorphous solids near a plastic event.  We show that the minimum eigenvalue distribution at a strain $\gamma$ near the plastic event occurring at $\gamma_{P}$, displays a collapse when scaled by $\sqrt{\gamma_P - \gamma}$, and with the number of particles as $N^{-0.22}$.  Notably, at low-frequencies, this scaled distribution displays a robust $D(\omega_{\min}) \sim \omega^{6}_{\min}$ power-law regime, which survives in the large $N$ limit. Finally, we probe the universal properties of this ensemble through a characterization of the second and third eigenvalues of the Hessian matrix near a plastic event.
\end{abstract}


\maketitle

\textit{Introduction:}
Amorphous solids are well known to display an anomalous temperature dependence
in their heat capacity~\cite{buchenau1991anharmonic, ramos2004calorimetric}.
This has been suggested to originate due to an excess of modes in their
vibrational density of states (VDoS), over and above the Debye modes of
crystalline systems, and is known as the Boson peak~\cite{buchenau1984neutron}.
This behavior is remarkably robust to the details of the models under
consideration, as well as the dimension of the system, and has emerged as a
hallmark of amorphous solids. Various theoretical models have been proposed in
order to reproduce and characterize this behavior~\cite{anderson1972anomalous,
phillips1972anomalous, buchenau2007neutron, baggioli2019universal,
zaccone2020relaxation, baggioli2020unified, casella2020physics,
baggioli2020new}.  Since a primary quantity of interest in the thermodynamic
limit are the mechanical properties of solid glasses, the relevant scales to
probe are their properties at low-temperatures, corresponding to
low-frequencies in the VDoS. Recently, a new vibrational characteristic of
glass formers has been identified: a regime displaying a $D(\omega) \sim
\omega^{4}$ scaling in the density of states~\cite{lerner2016vibration,
kapteijns2018nonphononic, paoluzzi2019relation, wang2019low,
arceri2020vibrational, richard2020universality, bonfanti2020universal,
shimada2020low, das2020robustness, paoluzzi2020probing, das2021universal}.
Many theoretical models built around two-level systems, replica symmetry
breaking, stress-correlations, random matrices and other hypotheses have been
proposed as the origin of this behavior~\cite{buchenau1991anharmonic,
gurevich2003anharmonicity, gurarie2003bosonic, parshin2007vibrational,
stanifer2018random, ikeda2019universal, cui2020vibrational,
bouchbinder2020lowfrequency, shimada2021novel, shimada2021vibrational,
conyuh2021random, corrado2021mean}, however the nature of the modes
contributing to the $\omega^4$ behavior are still a subject of active research.
In this context, it is important to characterize new, deviant universal
features and their connection to microscopic details.

One of the outstanding problems in the field of glass physics is the
development of a statistical, microscopic theory explaining their anomalous
thermodynamics. Despite considerable theoretical explorations, the best
understanding of the glassy regime of matter emerges from simulations.
Preparing an athermal, energy minimized ensemble of structural glass-formers
allows us to study an ensemble of \textit{rigid} configurations that are
amorphous in nature. Although assumed to be mechanically stable, such
configurations have been shown to contain an additional, strain degree of
freedom~\cite{dagois2012soft, wu2015hyperuniformity, *wu2015statistics,
*wu2015entropy}. An otherwise constrained configuration allows for unbalanced
shear stresses that may be specific to the simulation
parameters~\cite{footnote_stress}. The choice of appropriate stress ensembles
is then an important consideration in the study of amorphous
solids~\cite{henkes2009statistical, bi2013fluctuations}. While the effect of
modulating internal stresses have been studied~\cite{mizuno2017continuum,
lerner2018frustration, moriel2021internally}, it is pertinent to reexamine the
apparent universality~\cite{richard2020universality, bonfanti2020universal,
kapteijns2018nonphononic, shimada2020low} under physically relevant,
macroscopic shear stress ensembles.

In this Letter, we study realistic ensembles  of amorphous solids generated
through simple shear, most notably used in cyclic shearing protocols.  We use a
natural control parameter, namely the shear strain, in order to test the
sensitivity of the minimum eigenvalue distributions to changes in ensemble.  We
show that changes in the macroscopic shear-stress ($\sigma_{xy}$) which leave
the internal stress distributions invariant, results in a modification of the
amorphous VDoS from $D(\omega) \sim \omega^{4}$. We first consider a
shear-stabilized ensemble $(U = U_{\min}(\gamma))$ and show that the
low-frequency behavior of the VDoS shifts to a novel power-law close to
$D(\omega_{\min}) \sim \omega^{5}_{\min}$.  Such a constraint is relevant to
the study of stable solids which by definition, resist deformations.
Additionally, these results point to a link between the $\omega^{4}$ regime in
the VDoS and the stress fluctuations sustained by the system. We also uncover a
new universal distribution of the minimum eigenvalue under an ensemble of fixed
strain-distances to a plastic event. Notably, the distribution collapses under
suitable scaling of the strain as well as system size. This distribution
additionally displays a low-frequency behavior of $D(\omega_{\min}) \sim
\omega^{6}_{\min}$.

\begin{figure}[t!]
    \centering
    \resizebox{\linewidth}{!}{\input{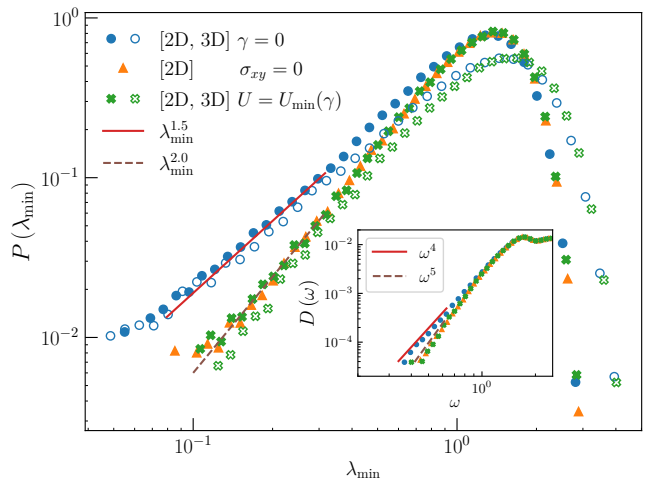}}
    \caption{Minimum eigenvalue distributions obtained from energy minimized
    configurations of a 2D system of $256$ particles. The unfilled markers
    correspond to a 3D system of $512$ particles. The plots compare a typical,
    \textit{Unstrained} ensemble (blue circles) against the shear-stabilized
    ensembles: \textit{Zero-shear-stress} (orange triangles) and
    \textit{Shear-strain-energy-minimized} (green crosses). The distributions
    drawn from these ensembles deviate significantly from the
    $\omega^{4}_{\min}$ regime.  The (red) solid and (violet) dashed lines
    correspond to power-laws of $\omega^{4}_{\min}$ and $\omega^{5}_{\min}$
    respectively. \textbf{(Inset)} Distribution of the \textit{full}
    vibrational density of states for a 2D system with $256$ particles. The low
    frequency behavior of the distribution is modified from $D(\omega) \sim
    \omega^{4}$ to $D(\omega) \sim \omega^{5}$.}\label{fig_isotropic}
\end{figure}

\textit{Minimum eigenvalue spectrum: }
Vibrational properties of a solid may be discerned from the Hessian of the
total potential energy $U\left[\{\mathbf{r}^{i j}\}\right] = \sum_{i j} \psi^{i
j}$, where $\psi^{i j}$ is the interaction potential between particles $i$ and
$j$ which we assume to be central. This is conveniently represented by the
Hessian matrix
\begin{equation}
    \mathcal{H}_{\alpha \beta}^{i j} (\mathbf{r}^{i j}) = \frac{\partial^{2} U
    \left[\{\mathbf{r}^{i}\}\right]}{\partial r^{i}_{\alpha} \partial
    r^{j}_{\beta}},
    \label{eq_hessian}
\end{equation}
the indexes of which run over dimensions $\alpha, \beta \in \{x,y,z\}$ for
every pair of particles $i,j \in \{1, \ldots , N\}$. Above, $r_{\alpha}^{ij}$
is the $\alpha$-component of the distance vector from particle $i$ to $j$. A
primary quantity of interest in the study of the vibrational properties of
glasses is the distribution of the \textit{minimum} eigenvalue of the Hessian
matrix, $\lambda_{\min}$.  This typically controls the longest time scales in
the system, and provides a useful route to characterize the stability of
amorphous solids~\cite{maloney2004universal}.

The vibrational frequencies are related to the eigenvalue of the Hessian as:
$\omega = \sqrt{\lambda}$. This allows us to relate the two distributions as
$D(\omega) = \sqrt{\lambda} P(\lambda)$. The distribution of $\lambda_{\min}$,
being an extreme value distribution is affected by the correlations between the
eigenvalues of the Hessian matrix. However, many glass formers display a
$P(\lambda_{\min}) \sim \lambda_{\min}^{1.5} \equiv D(\omega_{\min}) \sim
\omega_{\min}^4$ behavior in the tail of the minimum eigenvalue distribution,
indicating weak correlations in the low lying
eigenvalues~\cite{lerner2016vibration}.  Deviations from this universal
behavior are therefore of interest in determining different structural
properties of glasses.  Indeed, we show in this Letter that the response of
short ranged glass formers to shear is linked to changes in
$P(\lambda_{\min})$, which in turn is crucially sensitive to the
stress-ensemble from which configurations are drawn.

\begin{figure}[t!]
    \resizebox{\linewidth}{!}{\input{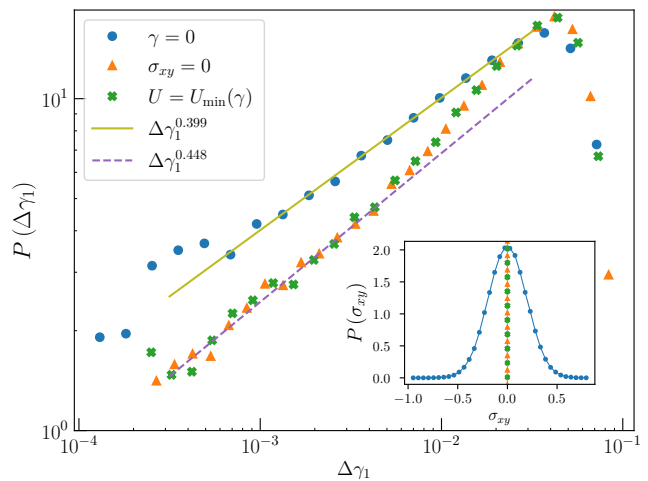}}
    \caption{Distributions of the strain $\Delta \gamma_1$ required to achieve
    the first plastic event, beginning with configurations sampled from the
    unstrained, and from the two shear-stabilized ensembles. These measurements
    were performed on 2D systems of size $N = 256$ in two dimensions.  The
    lines indicate best-fits for the exponent. The shear-stabilized ensembles
    have fewer plastic events at smaller strains.  \textbf{(Inset)}
    Distribution of the macroscopic shear stress in energy minimized
    configurations. The \textit{Unstrained} ensemble displays finite
    shear-stress fluctuations, while the shear-stabilized ensembles contain
    configurations with no macroscopic shear stress.}\label{fig_oneDrop2dR10}
\end{figure}

\textit{Shear-stabilized ensembles: }
We consider configurations that are allowed to undergo volume-preserving,
simple shear, where only the upper-triangular elements of the strain tensor can
be non-zero $(\gamma_{\alpha \beta} = \epsilon_{\alpha \beta}^{\alpha <
\beta})$. An isolated stable solid relaxes along all available degrees of
freedom. In such energy minimized configurations of systems comprised of
particles interacting via pairwise, central potentials, the off-diagonal
element of the macroscopic force moment tensor, i.e., shear stress is exactly
zero~\cite{karmakar2010athermal} (see Supplemental
Material~\cite{SI}/\ref{appendix_stable}):
\begin{equation}
    \frac{\partial U}{\partial \gamma_{\alpha \beta}} = \sum_{\langle i,j
    \rangle} f_{\alpha}^{ij} r_{\beta}^{ij}= \sum_{\langle i,j \rangle}
    \sigma_{\alpha \beta}^{ij} \equiv \sigma_{\alpha \beta} \times V,
    \label{eq_stress}
\end{equation}
where $f_{\alpha}^{ij}$ is the $\alpha$-component of the force on particle $i$
by particle $j$, $\sigma_{\alpha \beta}^{ij}$ is the bond-stress between
particles $i$ and $j$, $\sigma_{\alpha \beta}$ is the macroscopic stress tensor
and $V$ is the volume of the system. It is therefore natural to probe the
effect of macroscopic shear stress fluctuations on the stability properties of
such systems, which are enhanced due to relaxation along an additional strain
degree of freedom.

In this context, we analyze the distribution of minimum eigenvalues of the
Hessian matrices of configurations sampled from two different ensembles with
(i) finite shear stress fluctuations and (ii) zero shear stress fluctuations
(within a tolerance). The finite stress ensemble appears naturally when
generating energy minimized configurations from a thermal ensemble, under
periodic boundary conditions (see Supplemental
Material~\cite{SI}/\ref{appendix_stress}), and we refer to this as the
\textit{Unstrained} ensemble.  We create shear-stabilized ensembles using two
different procedures. In the first method, we athermally strain the system in
the direction of the shear stress until the stress changes sign, and this is
repeated two more times, with reduced strain increments. We term this ensemble
\textit{Zero-shear-stress} (Supplemental
Material~\cite{SI}/\ref{appendix_strain} shows the strains required). Such a
procedure allows us to attain stress-free states in systems with one
shear-stress, namely, systems in two dimensions. Therefore, we also use a
technique capable of relaxing stresses in three dimensional systems. In the
second method, we perform an energy minimization of the position as well as
shear strain degrees of freedom concomitantly, and refer to this ensemble as
\textit{Shear-strain-energy-minimized}.  Further details of both procedures are
described in the Supplemental Material~\cite{SI}/\ref{appendix_simulation}.
Notably, these protocols leave the statistics of the internal bond-stresses
invariant (see Supplemental Material~\cite{SI}/\ref{appendix_bond_stress}).

We display numerically sampled minimum eigenvalue distributions of the Hessian
for two (2D) and three dimensional (3D) systems in Fig.~\ref{fig_isotropic}.
Remarkably, the minimum eigenvalue distributions corresponding to the two types
of ensembles yield markedly different results, especially at the lowest
frequencies which govern large-scale stability properties. Specifically, we
find that the well-known $\omega^{4}_{\min}$ regime is modified in the
shear-stabilized ensembles, and instead we find the best-fit power-law to be
closer to $\omega^{5}_{\min}$. For data on larger system sizes, see
Supplemental Material~\cite{SI}/\ref{appendix_large}. Moreover, we find that
the two different procedures of generating a shear-stabilized ensemble yield
identical distributions, pointing to the fact that these distributions are
sensitive to the stress ensemble and not the preparation protocol, independent
of dimension.

\textit{Mechanical properties: }
Understanding the relationship between microscopic parameters and bulk rigidity
is important in constructing a first-principles theory of solids.  In order to
further probe the connection between the minimum eigenvalue distributions and
the stability of configurations created in different stress ensembles, we carry
out Athermal Quasistatic Shearing (AQS) of the
system~\cite{kobayashi1980computer}, using $2\text{D}$ glass structures (see
Supplemental Material~\cite{SI}/\ref{appendix_aqs} for details).  AQS allows us
to trace the state of a local minimum as the potential energy surface is
transformed under an effectively infinitesimal strain rate. Amorphous materials
as well as crystals, when subjected to an incremental strain, produce a
corresponding linear stress-response. However, unlike crystals, amorphous
arrangements of particles incur localized, non-affine, displacements termed
`plastic events'. These deformations are easily identified in an athermal
straining protocol by the occurrence of abrupt stress-drops and localized
particle displacements. The amorphous nature of the constituent particles
allows the system to release stresses via such events that comprise
displacements of a small fraction of the particles that occurs when energy
minimizing the system after imparting it an affine strain.

\begin{figure}[t!]
    \resizebox{\linewidth}{!}{\input{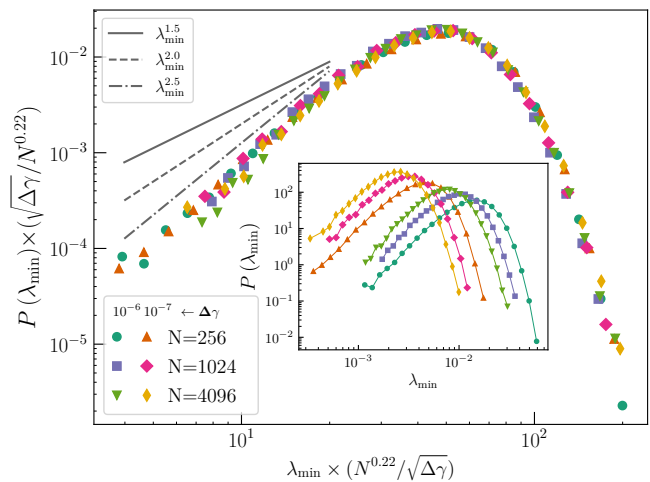}}
    \caption{Distributions of the minimum eigenvalue of the Hessian matrix,
    $\lambda_{\min}$, drawn from strained configurations grouped by distance
    $\Delta \gamma$ from their respective first plastic events
    \textbf{(Inset)}.  These distributions collapse when scaled by the strain
    gap as $\sqrt{\Delta \gamma}$ and the number of particles as $N^{-0.22}$.
    These scaled distributions show a marked deviation from $\omega_{\min}^4$
    behavior (solid line). The dash-dotted line corresponds to a power-law of
    $\omega_{\min}^6$.}\label{fig_universal}
\end{figure}

The distribution of the strain needed to induce the first plastic event forms
an important descriptor of the rigidity of solids, and is an indicator of their
stability to shear. It is therefore important to study the nature of such
distributions in appropriate, experimentally relevant stress-ensembles.  As
discussed in Eq.~(\ref{eq_stress}), the shear-stabilized ensembles with zero
shear stress may provide an accurate characterization of the stability of real
solids. In Fig.~\ref{fig_oneDrop2dR10} we show that the distribution of the
strain $\Delta \gamma_{1}$ needed to achieve the first plastic event, is
sensitive to the stress-fluctuations allowed in the configurations sampled.
Most significantly, the \textit{Unstrained} ensemble is more susceptible to
plastic events at lower strain-deformations. Intriguingly, the estimated
exponent $(P(\Delta \gamma_{1}) \sim {\Delta \gamma_{1}}^{\theta})$ in the low
$\Delta \gamma_{1}$ regime, an important characterization of amorphous
stability~\cite{lin2014density, lin2014scaling}, seems to increase from $\theta
\approx 0.4$ to $\theta \approx 0.45$ between the unstrained and
shear-stabilized ensembles.

\textit{Plastic-event approach ensemble: }
Since a primary utility of a Hessian analysis is the determination of the
stability of amorphous systems, it is natural to focus on the nature of the
ensemble of near-failure amorphous solids. These plastic events correspond to
the system crossing saddles in the energy landscape as it is
sheared~\cite{maloney2006aqs}. Traversing across such energy barriers by
straining the system allows us to probe the energy landscape that determines
the stability of such amorphous configurations of particles. The model system
used allows us to study its properties close to such a phenomenon. Once the
plastic event is identified, as described in the previous section, we then
proceed to ascertain the strain $\gamma_{P}$, at which the plastic event
occurs, to a high degree of precision by using very fine strain-steps. This
permits us to sample configurations that are arbitrarily close to the event. We
thus define the \textit{Plastic-event-approach} ensemble as a collection of
configurations all at the same strain to the plastic event $(\Delta\gamma =
\gamma_{P} - \gamma)$.

We study the single most important marker of stability, namely the minimum
eigenvalue of the Hessian, as the system approaches the plastic-strain
$(\gamma_{P})$ at which a saddle in the energy landscape is reached.  The
behavior of the displacement field has been shown to be proportional to the
minimum eigenmode, when close to such a plastic
event~\cite{karmakar2010predicting}:
\begin{equation}
    \mathbf{u}(\gamma) - \mathbf{u}(\gamma_{P}) = X(\gamma) \mathbf{\Psi}_{\min},
\end{equation}
where $\mathbf{u}$ represents the position of the particles as a function of
the strain $\gamma$, and $X$ is the projection of the displacement field on to
the minimum eigenvector $\mathbf{\Psi}_{\min}$. The minimum eigenvalue is
assumed to vary linearly with the projection $\lambda_{\min} \approx \alpha
X(\gamma)$, which in-turn leads to an approach to zero with a square-root
singularity: $\lambda_{\min} \approx \alpha \sqrt{\gamma_{P} - \gamma}$.  This
singular behavior occurs due to the eigenvector corresponding to the minimum
eigenvalue aligning itself with the displacement vector corresponding to the
plastic event. A natural question then is the exact nature of the
proportionality constant $\alpha$ that governs the magnitude of the change in
the minimum eigenvalue of the Hessian with the strain of the system. The
singular square-root approach is quite general and is expected whenever a
system approaches a saddle corresponding to a plastic event along one of its
degrees of freedom. For example, a crystalline system undergoing a slip will
have its eigenvalue vanish with a single $\alpha$ determined by the
interactions between the particles. On the other hand, amorphous materials
differ in that the constant of proportionality $\alpha$ varies from sample to
sample. The statistics of $\alpha$ is consequently dependent purely on the
microscopic parameters of the system, and we therefore expect a universal
distribution of the form
\begin{equation}
    P(\alpha) \equiv P \left(\frac{\lambda_{\min}}{\sqrt{\gamma_{P} - \gamma}}\right).
\end{equation}
In Fig.~\ref{fig_universal}, we display these distributions at small distances
to the plastic strain as well as for various system sizes. We scale these
distributions with the strain-distance as $\sqrt{\Delta\gamma}$ and system size
as approximately $N^{-0.22}$. This universal distribution seems to exhibit a
low-frequency power-law of $\lambda^{2.5}_{\min}$ corresponding to
$\omega^{6}_{\min}$. The full approach to the plastic  event is illustrated in
the Supplemental Material~\cite{SI}/\ref{appendix_approach}.  Attempts at
fitting one of the three common extreme value distributions failed to yield a
reasonable match, suggesting a non-trivial limiting form.
Using the estimated exponent $\alpha \approx 2.5$  in an extreme value fit of
uncorrelated variables predicts a scaling with $N$ with an exponent
$1/(1+\alpha) \approx 0.286$. The significant difference from our observed
system size scaling exponent of $0.22$ also points to correlations in the
underlying eigenvalues, which would be interesting to characterize further.

\begin{figure}[t!]
    \resizebox{\linewidth}{!}{\input{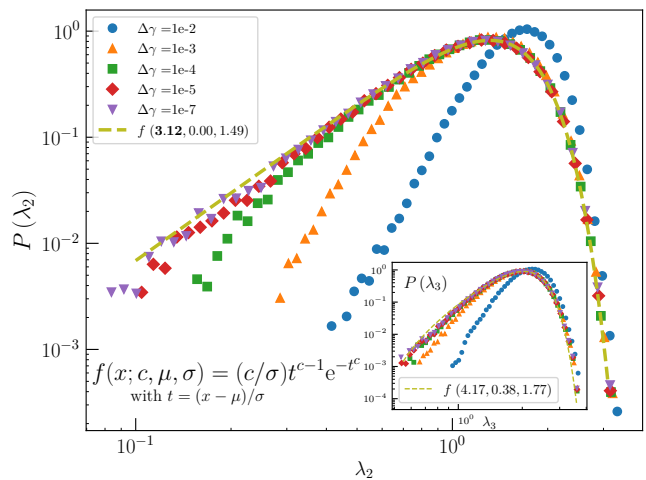}}
    \caption{Distributions of the second eigenvalue of $N = 256$ systems, as it
    approaches the plastic event. Surprisingly, the distribution approaches a
    Weibull form as the plastic event is approached. The dashed line is a
    two-parameter Weibull fit with $\mu$ set to $0$. \textbf{(Inset)}
    Distribution of the third eigenvalue in the same system. This distribution
    does not fit well to a generalized extreme value form. The best fit is
    displayed by the dashed line.}\label{fig_lvlSpace}
\end{figure}

\textit{Second and third eigenvalue distributions: }
The Hessian matrices of amorphous systems have also been sought to be modeled
within Random Matrix frameworks~\cite{stanifer2018random, beltukov2011sparse,
beltukov2013ioffe, manning2015peak, conyuh2017bosonpeak,
baggioli2019vibrational}.  In this context, we analyze the behavior of the
second and third eigenvalues $\lambda_{2}, \lambda_{3}$, as the system
approaches a plastic event. Notably, in the limit of a vanishing minimum
eigenvalue, the second eigenvalue is equivalent to the first level-spacing.
Such near-extreme value distributions are natural measures that arise in Random
Matrix Theory~\cite{mehta2014random, porter1965statistical}, and could
therefore serve as useful tools to understand the nature of the
\textit{ensemble} that the Hessian matrices of amorphous solids generate.

As the plastic event is approached, the minimum eigenvalue departs from the
remaining vibrational frequencies. The effect of such a separation is clearly
felt by the remnant of the spectrum, as can be seen in Fig.~\ref{fig_lvlSpace}.
Interestingly, as the system approaches this saddle point, the distribution of
the second eigenvalue converges to a zero-located Weibull distribution. Such
Weibull forms have also been observed in the minimum eigenvalue distributions
in glass formers, for small system sizes~\cite{lerner2016vibration}.  Our best
fit curve is displayed in Fig.~\ref{fig_lvlSpace}, showing a very good match.
Additionally, the fit estimates a low-frequency power-law of about
$\lambda^{2.12}_{2}$. Such a characterization assumes relevance when studying
solids close to plastic events, because the VDoS may then be well represented
by a spectrum with one less mode than otherwise. Finally we also measure the
statistics of the third eigenvalue as the plastic event is approached. We plot
this distribution in the inset of Fig.~\ref{fig_lvlSpace}.  Once again, this
distribution attains a limiting form. However, this distribution does not seem
to fit well with the generalized extreme value distributions.

\textit{Discussion: }
We have presented results highlighting the role played by the choice of
ensemble in the low-frequency regime of the VDoS of structural glass formers.
We find that crucially, finite shear stress fluctuations are required to
observe the universal $\omega_{\min}^4$ regime that has emerged as a hallmark
of low-temperature glasses. Determining the appropriate distributions of
stresses in real amorphous solids prepared under different conditions, and
their effect on structural properties would therefore be of immediate
relevance.  We also showed that the minimum eigenvalue of the Hessian attains a
universal distribution when approaching a plastic event. It would be
interesting to probe the origin of the anomalous scaling of $N^{-0.22}$ with
the number of particles displayed by this distribution. The robustness of the
$\omega_{\min}^4$ regime in the VDoS of amorphous solids in the context of our
study motivates an analysis of different models of structural glass formers in
stress-controlled ensembles, in two as well as three dimensions.  Similarly,
studying the effects of varying the smoothness in the interaction potentials
which have been shown to have non-trivial effects on the Hessian
matrices~\cite{krishnan2020singularities}, could help better understand the
stability of amorphous solids to shear.  Finally, it would also be interesting
to study the shear stress fluctuations in ultrastable glasses, which have been
shown to have anomalous rigidity properties~\cite{ozawa2018random}.

\textit{Acknowledgments: }
We thank Edan Lerner\thinspace\orcidicon{0000-0001-9245-6889}, Jishnu Nampoothiri\thinspace\orcidicon{0000-0001-9801-5072} and Srikanth Sastry\thinspace\orcidicon{0000-0001-7399-1835} for useful discussions. V.V.K.~thanks the Council of Scientific and Industrial Research, India for support via the Shyama Prasad Mukherjee Fellowship (SPM-07/1142(0228)/2015-EMR-1). S.K.~would like to acknowledge the support from Swarna Jayanti Fellowship Grants No.~DST/SJF/PSA-01/2018-19 and No.~SB/SFJ/2019-20/05.  This project was funded by intramural funds at TIFR Hyderabad from the Department of Atomic Energy, Government of India.

\bibliographystyle{apsrev4-2} 
\bibliography{PlasticApproach_Bibliography}

\clearpage

\begin{widetext}

\begin{appendix}
\renewcommand{\thefigure}{S\arabic{figure}}
\setcounter{figure}{0}

\section*{\large \texorpdfstring{S\MakeLowercase{upplemental}
M\MakeLowercase{aterial for}\\ ``U\MakeLowercase{niversal
non-}D\MakeLowercase{ebye low-frequency vibrations in sheared amorphous
solids}''} {Supplemental Material for "Universal non-Debye low-frequency
vibrations in sheared amorphous solids"}}

\subsection{Shear-stabilization}~\label{appendix_stable}

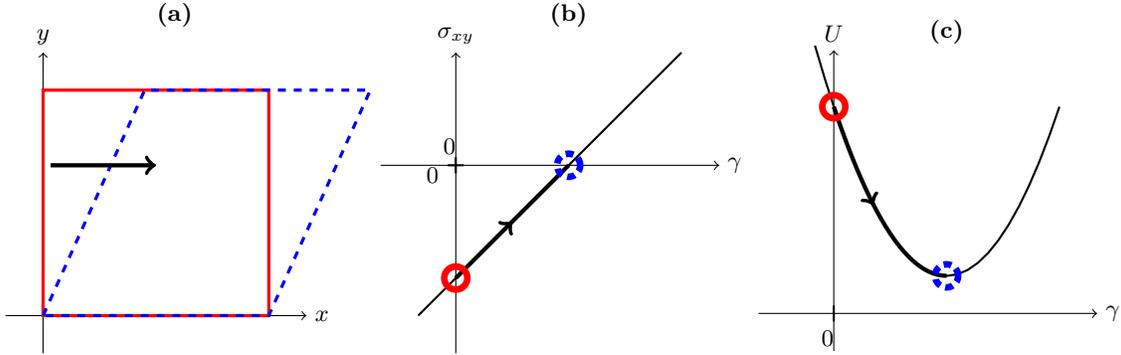
\begin{figure}[ht]
    \centering
    \usetikzlibrary{decorations.markings}
    \begin{tikzpicture}
        \draw[->, thin] (-0.5,0) -- (3.5,0) node[right] {$x$};
        \draw[->, thin] (0,-0.5) -- (0,3.5) node[above] {$y$};
        \draw[very thick, red] (0, 0) rectangle (3.0, 3.0);
        \draw[very thick, dashed, xslant=0.45, blue] (0, 0) rectangle (3.0,3.0);
        \draw[->, ultra thick] (0.1,2.0) -- (1.5,2.0);
        \draw (1.75,4.0) node[] {\textbf{(a)}};
    \end{tikzpicture}
    \begin{tikzpicture}[domain=-0.5:3]
        \draw[->, thin] (-1.0,0) -- (3.5,0) node[right] {$\gamma$};
        \draw[->, thin] (0,-2.5) -- (0,1.5) node[above] {$\sigma_{xy}$};
        \draw[thick] (0,3pt) -- (0,-3pt) node[above,xshift=-0.6ex,yshift=1ex] {$0$};
        \draw[thick] (3pt,0) -- (-3pt,0) node[left,yshift=-1ex] {$0$};
        \draw[thick] plot (\x, {(\x - 1.5)});
        \begin{scope}[ultra thick,decoration={markings,mark=at position 0.5 with {\arrow{>}}}]
        \draw[postaction={decorate}] (0, -1.5) -- (1.5, 0);
        \end{scope}
        \draw (0, -1.5) node[circle,thick,red,draw,line width=2.5pt] {};
        \draw (1.5, 0) node[circle,thick,densely dashed,blue,draw,line width=2.5pt] {};
        \draw (1.5,2.0) node[] {\textbf{(b)}};
    \end{tikzpicture}
    \begin{tikzpicture}[domain=-0.25:3]
        \draw[->, thin] (-1.0,0) -- (3.5,0) node[right] {$\gamma$};
        \draw[->, thin] (0,-0.5) -- (0,3.5) node[above] {$U$};
        \draw[thick] (0,3pt) -- (0,-3pt) node[below,xshift=-0.6ex] {$0$};
        \draw[thick] plot (\x, {(\x - 1.5)^2 + 0.5});
        \begin{scope}[ultra thick,decoration={markings,mark=at position 0.5 with {\arrow{>}}}]
        \draw[domain=0:1.5, postaction={decorate}] plot (\x, {(\x - 1.5)^2 + 0.5});
        \end{scope}
        \draw (0, 2.75) node[circle,thick,red,draw,line width=2.5pt] {};
        \draw (1.5, 0.5) node[circle,thick,densely dashed,blue,draw,line width=2.5pt] {};
        \draw (1.5,3.75) node[] {\textbf{(c)}};
    \end{tikzpicture}
    \caption{\textbf{(a)} Schematic representations of a system undergoing
    simple shear, and corresponding changes in \textbf{(b)} stress and
    \textbf{(c)} energy. The (red) solid state represents an unstrained state
    that exhibits a finite shear-stress. The (blue) dashed state represents a
    shear-stabilized state.}\label{fig_schematic}
\end{figure}

\subsection{Stress Fluctuations}~\label{appendix_stress}

\begin{figure}[ht]
    \resizebox{0.6\linewidth}{!}{\input{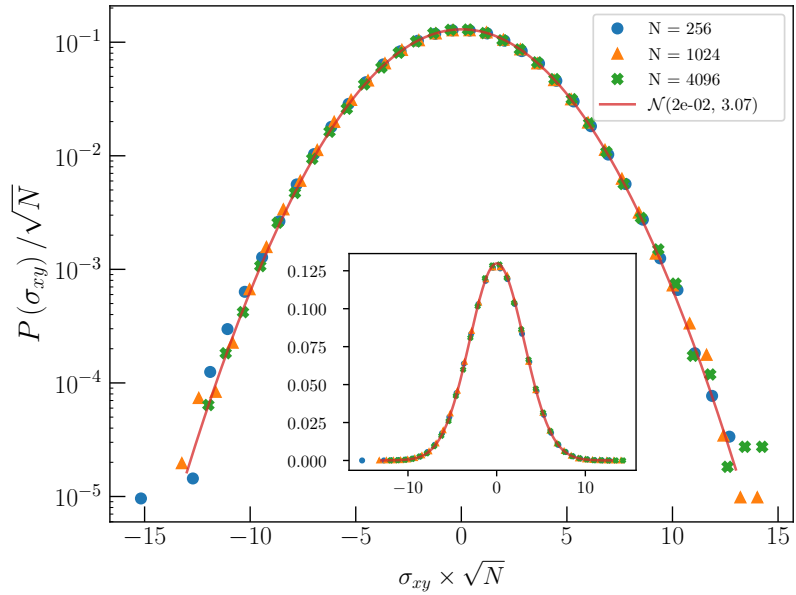}}
    \caption{Stress distributions of configurations generated by cooling and
    energy-minimizing a thermal ensemble under periodic boundaries. The plot
    shows the stress fluctuations of system sizes $N \in \{256, 1024, 4096\}$.
    The distributions scale with system size as $1/\sqrt{N}$ in two dimensions.
    The solid line is a maximum-likelihood-estimate fit of the Normal
    distribution to the data corresponding to $N = 4096$, with fit parameters
    shown in the legend. These plots quantify the effective residual stresses
    present in simulated models of amorphous solids when prepared under
    unstrained conditions.}\label{fig_stress}
\end{figure}

\clearpage
\subsection{Strain required to achieve Shear-stabilization}~\label{appendix_strain}

\begin{figure}[ht]
    \resizebox{0.47\linewidth}{!}{\input{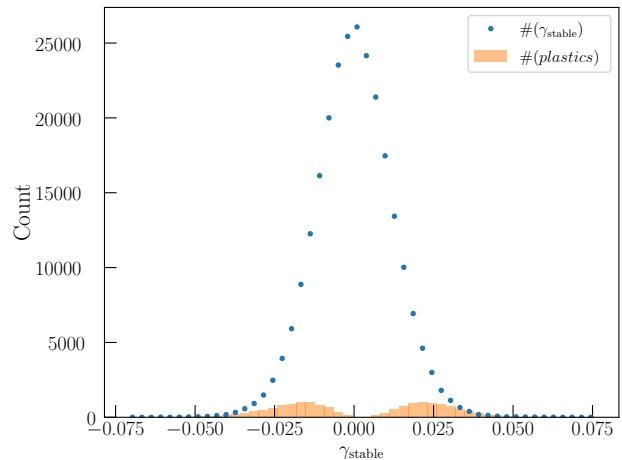}}
    \caption{Frequency distribution of strains required to attain
    shear-stability, in the \textit{Zero-shear-stress} protocol, in systems of
    size $N = 256$. The orange bars indicate the number of plastic events
    encountered by samples that undergo strains in the corresponding intervals.
    They account for $\sim 6.2\%$ of all samples. In the
    \textit{Zero-shear-stress} protocol, we encounter plastic events during
    AQS\@. Here we show that the proportion of straining trajectories that
    encounter them, while small, is not insignificant.}\label{fig_strain}
\end{figure}

\subsection{Internal Bond-Stresses}~\label{appendix_bond_stress}

\subsubsection{Distributions}~\label{appendix_bond_dist}

\begin{figure}[htb]
    \resizebox{\linewidth}{!}{\input{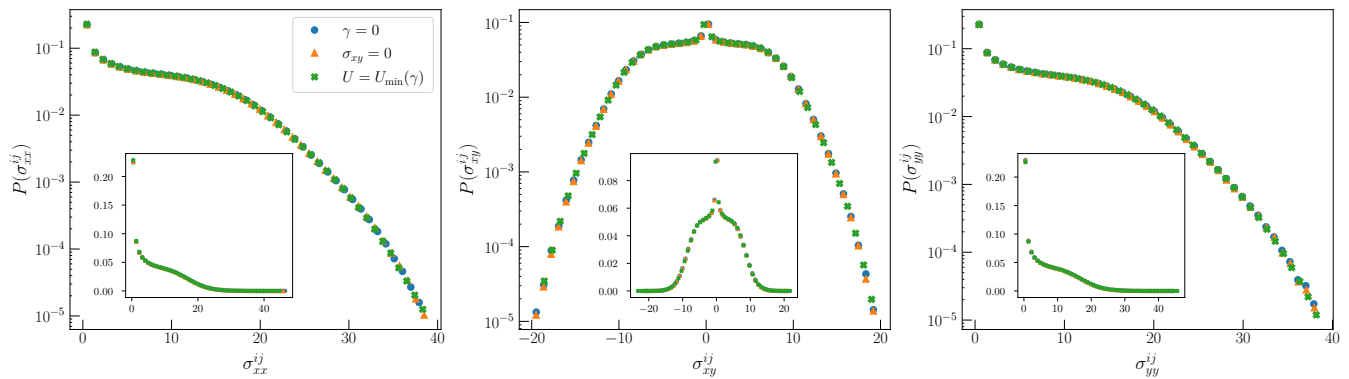}}
    \caption{Distributions of bond-stresses, from an ensemble of configurations
    of systems of size $N=256$. The \textit{Unstrained},
    \textit{Zero-shear-stress} and \textit{Shear-strain-energy-minimized}
    ensembles all exhibit \textit{identical}
    distributions.}\label{fig_bondStress_dist}
\end{figure}

\subsubsection{Visualization}~\label{appendix_bond_config}


\begin{figure}[hp]
    \resizebox{0.85\linewidth}{!}{\input{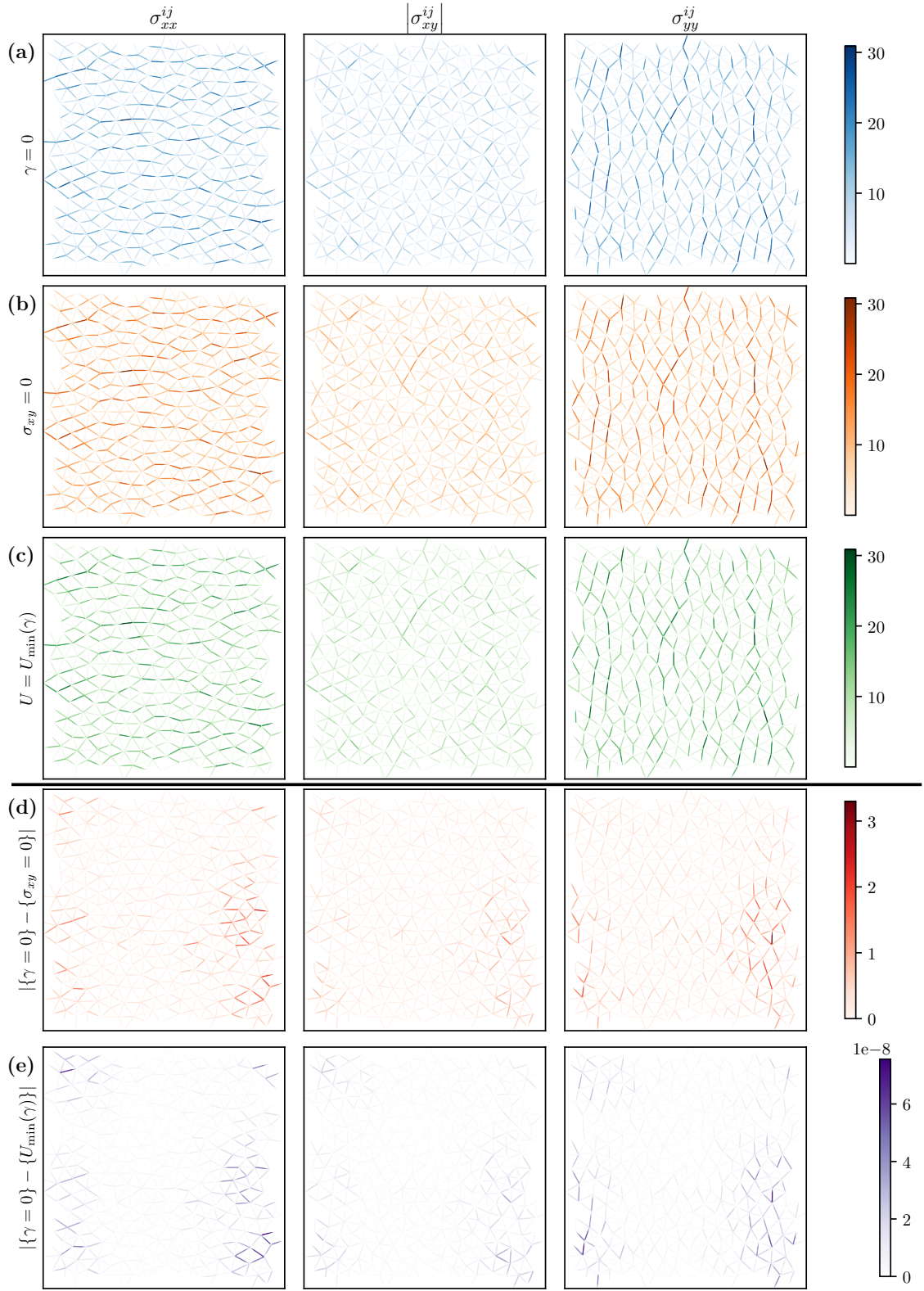}}
    \caption{Components of the stress tensor at each `bond' in a
    two-dimensional configuration of a system of size $N=256$. \textbf{(a)} an
    unstrained configuration, \textbf{(b)} The same configuration strained to
    achieve zero-shear-stress. \textbf{(c)} The same configuration as in
    \textbf{(a)}, energy minimized with a strain degree of freedom.
    \textbf{(d)} The difference in bond stresses between the
    \textit{Unstrained} and \textit{Zero-shear-stress} configurations is an
    order of magnitude smaller than the original stress. \textbf{(e)} The
    difference in bond stresses between the \textit{Zero-shear-stress} and
    \textit{Shear-strain-energy-minimized} configurations is zero up to
    numerical precision.}\label{fig_bondStress_conf}
\end{figure}

\clearpage
\subsection{Large Systems}~\label{appendix_large}

\begin{figure}[ht]
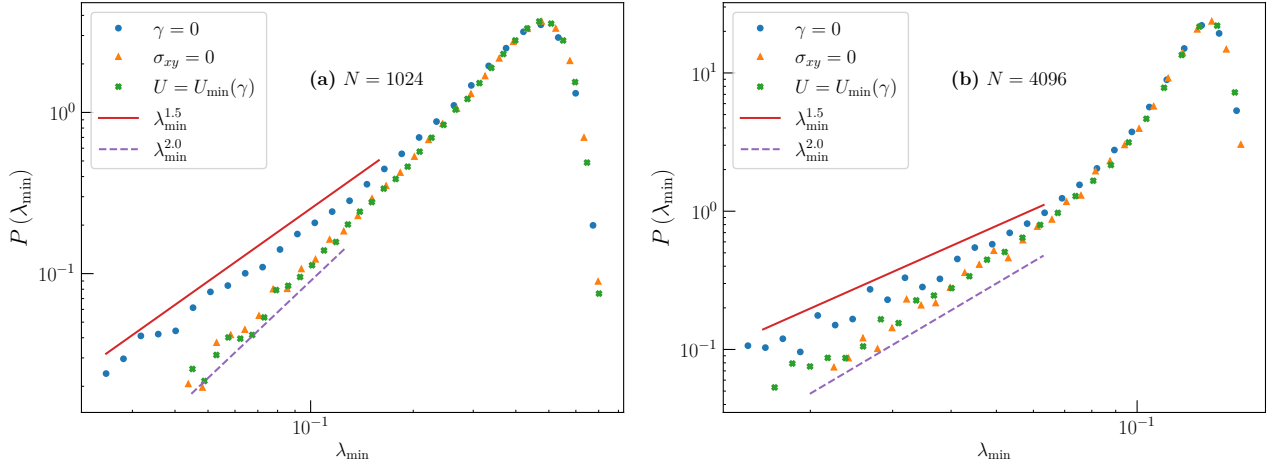

    \resizebox{0.47\linewidth}{!}{\input{minEigenIsotropic_2dR10_T-058_smooth2_01024.pgf}}
    \resizebox{0.47\linewidth}{!}{\input{minEigenIsotropic_2dR10_T-058_smooth2_04096.pgf}}
    \caption{Minimum eigenvalue distributions of the \textit{Unstrained} and
    shear-stabilized ensembles in 2D, corresponding to larger system sizes: (a)
    $N = 1024$ and (b) $4096$. The lines indicate pure power-laws. The
    deviation from $\omega^{4}_{\min}$ persists as larger system sizes are
    probed, and is not diminished with the suppressed stress fluctuations shown
    in Fig.~\ref{fig_stress}.}\label{fig_larger}
\end{figure}

\subsection{Approach to Universal Distribution}~\label{appendix_approach}

\begin{figure}[hb]
    \resizebox{\linewidth}{!}{\input{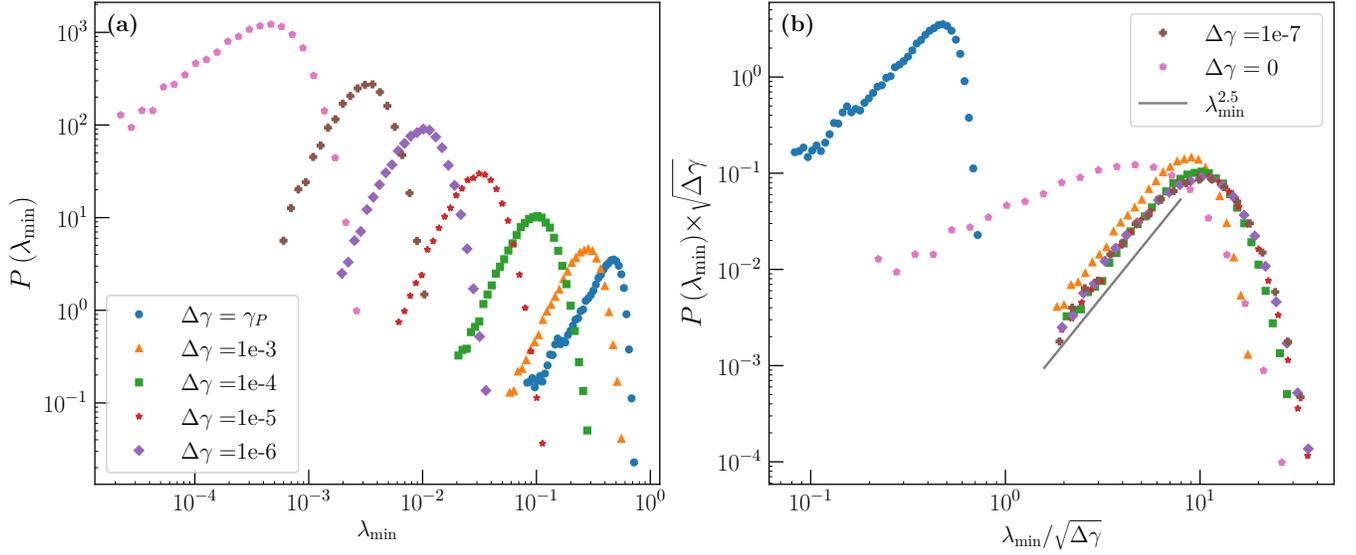}}
    \caption{Minimum eigenvalue distributions of a system of size $N = 1024$,
    upon approaching a plastic event. The (blue) circles display the
    distribution of eigenvalues in the initial, unstrained state. The (pink)
    hexagons are distributions of eigenvalues measured at closest approach to
    the plastic strain $\gamma_{P}$. The other distributions belong to
    ensembles of structures at specific strains away from
    $\gamma_{P}$.}\label{fig_approach}
\end{figure}

%

\subsection{Additional Simulation Details}
\label{appendix_simulation} 

\subsubsection*{Simulation Potentials}

We simulate a 50:50 mixture of two particle types A and B. The interaction
potentials are cut-off at a distance
\begin{equation}
    r_{c} = 1.385418025 \; \sigma,
\end{equation}
with the three interaction diameters given by
\begin{align}
    \sigma_{A A} &= 1.0, \nonumber \\
    \sigma_{B B} &= 1.4, \nonumber \\
    \sigma_{A B} &= \sqrt{\sigma_{A A}\sigma_{B B}}.
\end{align}
The only difference between the parameters in two and three dimensions of
this model are the reduced number-densities given by
\begin{align}
    \rho_{2\text{D}} &= 0.85, \nonumber \\
    \rho_{3\text{D}} &= 0.81.
\end{align}

In our simulations we focus on the purely repulsive pairwise potential, given
by a tenth order polynomial, termed `R10'. The potential smooth to $n$
derivatives at cut-off is given by
\begin{align}
    \psi &= {\left(\frac{\sigma}{r}\right)}^{10} + \sum_{m=0}^{n} c_{2 m} {\left(\frac{r}{\sigma}\right)}^{2 m}
\end{align}
where the constants are calculated appropriately. We use only even-powered
polynomials in order to avoid the potential curving downwards at the cutoff, to
any precision, thus eliminating any attraction at the cutoff.

\subsubsection*{Sample size}
\begin{table}[h]
    \centering
    \caption{Number of minimum eigenvalue samples collected toward binning the
    $P(\lambda_{\min})$ histograms plotted in Figs.~\ref{fig_isotropic}
    and~\ref{fig_larger}. The suffix `K' indicates a
    thousand.}\label{tab_samples1}
    \begin{ruledtabular}
    \begin{tabular}{@{}lcccc@{}}
        Dimension               & \multicolumn{3}{c}{2} & 3  \\
        \cmidrule(lr){2-4}
        System size $(N)$       & 256 & 1,024 & 4,096 & 512 \\
        Samples                 & 256K & 256K & 150K & 256K
    \end{tabular}
    \end{ruledtabular}
\end{table}

\begin{table}[h]
    \centering
    \caption{Number of minimum eigenvalue samples collected toward binning the
    \textit{Plastic-event approach} $P(\lambda_{\min,2,3})$ histograms plotted
    in Figs.~\ref{fig_universal},~\ref{fig_lvlSpace} and~\ref{fig_approach}.
    The suffix `K' indicates a thousand.}\label{tab_samples2}
    \begin{ruledtabular}
    \begin{tabular}{@{}lccc@{}}
        Dimension               & \multicolumn{3}{c}{2} \\
        \cmidrule(lr){2-4}
        System size $(N)$       & 256 & 1,024 & 4,096 \\
        Samples                 & 256K & 50K & 50K
    \end{tabular}
    \end{ruledtabular}
\end{table}

\subsubsection*{Software}
Simulations of glasses along with the energy minimizations were performed using
LAMMPS~\cite{plimpton1995fast, lammps}. The stopping criterion for the
minimization was the force 2-norm: $\sqrt{\sum_{i=1}^{N} {\lvert F_{i}
\rvert}^{2}}$.  Eigenvalue calculations were performed using the
LAPACK~\cite{laug} routine: \texttt{dsyevr}, for small systems, and the Intel
MKL~\cite{intelmkl} sparse solver routine: \texttt{mkl\_sparse\_d\_ev}, for
large-sized matrices.  Analyses were performed with the help of
NumPy~\cite{2011numpy, harris2020array, numpy} and SciPy~\cite{2020SciPy,
scipy} libraries. Plotting was performed using
Matplotlib~\cite{hunter2007matplotlib, matplotlib}.

\subsubsection*{Athermal Ensembles}
\label{appendix_aqs} 

\textit{Unstrained ensemble:}
In our simulations, we use two-dimensional glass formers with varying particle
numbers $N \in \{256, 1024, 4096\}$ and a three-dimensional system of size $N =
512$, equilibrated at a parent temperature $T_{p}=0.58$. We then cool to
near-zero temperature at a slow rate of $\dot{T} \approx 10^{-2}$. We then
employ the conjugate gradient algorithm to achieve an energy minimized state up
to a force tolerance of $1.0 \times 10^{-10}$. These configurations constitute
the \textit{Unstrained} ensemble.  We also use these configurations to generate
the shear-stabilized ensembles.

\textit{Zero-shear-stress ensemble:}
We begin with a configuration drawn from the \textit{Unstrained} ensemble and
calculate the total shear stress $(\sigma_{\text{xy}})$. We then strain the
configuration in the direction of the stress. For example, if the shear stress
is negative, then the system is strained towards the \textit{left}. This choice
of straining direction is determined by the direction of the initial stress in
each configuration. This causes the stress to decrease in magnitude, and we
proceed until the stress reverses direction. We perform the same operation two
more times, each time with decreasing strain increments. The three strain steps
we use are: $\Delta\gamma \in \{5 \times 10^{-5}, 10^{-8}, 10^{-11}\}$. The
eigenvalues typically were not seen to vary much beyond the first strain step,
but we proceed to ensure that we are not separated from the shear-stabilized
state by a plastic event. When performing Athermal Quasi-static Shearing (AQS),
we use Lees-Edwards boundary conditions, and strain at an engineering strain
rate of $5.0 \times 10^{-5}$. At every step, the structure is relaxed to its
minimum energy, to a force tolerance of $1.0 \times 10^{-10}$.

\textit{Shear-strain-energy-minimized ensemble:}
The configurations were generated using the LAMMPS procedure
\texttt{box/relax}. The primary utility of this algorithm is that it allows one
to perform energy minimizations allowing the shape of the simulation box to
change, while also maintaining periodic boundaries. We make use of the
procedure with only the shear-strain included as a degree of freedom aside from
the particle positions. As highlighted in the documentation~\cite{lammps}, this
method encounters issues due to the algorithm utilizing the initial,
un-strained box dimensions as a reference for the stress computation. The
effect of this is that configurations that are at a large strain away from an
stable state as well as those configurations that suffer plastic events before
attaining shear-stability, both fail to achieve minimization to the desired
force tolerance. This is remedied, as suggested in the documentation, in two
ways: first by utilizing the \texttt{nreset} option to recalculate the
reference box dimensions, and second by restarting the minimizer upon failure,
typically across plastic events.

\textit{Plastic-event-approach ensemble:}
Plastic events are said to have occurred when there are non-affine
displacements with a localized spatial extent and a small fraction of
participating particles. These displacements differ from the typical elastic,
affine response of the particles to the applied strain, most significantly in
that the total magnitude of the displacement are much larger. An important
feature of these events is the quadrupolar nature of the displacement field,
centered at the point of localization, signaling a \textit{T1}-like event. In
order to `detect' a plastic event, we utilize a convenience of the AQS
protocol, being that every step of straining involves two stages: (a)
application of an affine strain and (b) an energy minimization. Plastic events
present large displacements in stage (b) of the protocol. Therefore, we keep
track of the displacement of the maximally displaced particle at every step of
energy minimization, and we register a plastic event when that value crosses a
threshold. For our model, we use a value of $\approx 15 \times \delta\gamma$,
whereas under elastic conditions, the maximum displacements during minimization
are $\sim \delta\gamma$. Additionally in order to avoid some corner-case
scenarios, we also utilize a minimum energy threshold of $10^{-9}$ energy units
for a step to register as a plastic event at a particular strain $(\gamma_{P})$.

Given such a mechanism to detect plastic events, we now define an ensemble of
configurations that all need the same `strain' to incur a plastic event
$(\Delta\gamma = \gamma_{P} - \gamma)$. Note that $\Delta\gamma > 0$. In order
to sample small enough values of $\Delta\gamma$, we first measure the
plastic-strain, $\gamma_{P}$ to an accuracy of $10^{-8}$ by `back-tracking' to
a previous state upon encountering a plastic-event, and subsequently straining
the system at the requisite precision. Thus, we are able to sample
configurations of the system at various values of the strain-to-plastic-event:
$\Delta\gamma \in \{10^{-3}, 10^{-4}, 10^{-5}, 10^{-6}, 10^{-7}\}$.

\end{appendix}

\newpage
\end{widetext}

\end{document}